\documentclass{elsart}
\usepackage{graphicx}
\usepackage{txfonts}

\begin{document}

\begin{frontmatter}

\title{Stability of Orbits around a Spinning Body in a Pseudo-Newtonian Hill Problem}

\thanks[mailsteklain]{steklain@ime.unicamp.br}
\thanks[mailletelier]{letelier@ime.unicamp.br}

\author{A. F. Steklain\thanksref{mailsteklain}}
\author{P. S. Letelier\thanksref{mailletelier}}

\address{Departamento de Matem\'atica Aplicada, Instituto de Matem\'atica, Estat\'{\i}stica e Computa\c{c}\~ao Cient\'{\i}fica,
Universidade Estadual de Campinas, 13083-970, Campinas, S\~ao Paulo, Brazil}

\begin{abstract}
A pseudo-Newtonian Hill problem based on a potential proposed by Artemova et al. [Astroph. J. 461 (1996) 565] is presented. 
This potential reproduces some of the general relativistic effects due to the spin angular momentum of the bodies, like the dragging of 
inertial frames.  Poincar\'e maps, Lyapunov exponents and fractal escape techniques are employed to study the stability of bounded and 
unbounded orbits for different spins of the central body.
\end{abstract}


\end{frontmatter}

\section{Introduction}

The Hill Problem was first formulated by Hill \cite{Hill}, in order to study the Moon-Earth-Sun system. This is a special case of the 
circular, planar restricted three-body problem, as described by Murray and Dermott \cite{Murray} and Arnold \cite{Arnold}, where the movement 
of the 
Moon around the Earth was just perturbed by a distant Sun. The Hill Problem is still applied in solar system models where bodies in nearly 
circular orbits are perturbed by other far away  massive bodies, and is very useful in the study of the stellar
dynamics. In many systems the Hill problem can be taken as a first approximation and can easily accommodate necessary modifications (see 
for instance Heggie \cite{Heggie}). The interaction of a Keplerian binary system with a normally incident circularly polarized 
gravitational wave can be represented by a Hill system, as shown by Chicone et al. \cite{Chicone}. The Hill problem was proved to be 
non-integrable by Meletlidou et al. \cite{Meletlidou}, and is chaotic, as shown by Sim\'o and Stuchi \cite{Stuchi}. 

The Hill Problem was first formulated in the context of the Newtonian dynamics.   In this dynamics  particles or spheres with or without  
spin have the same gravitational potential. However, on the 
General Relativity rotation modifies the metric and create effects like the dragging of inertial frames, or Lense-Thirring effect 
\cite{LenseThirring}. In General Relativity the exterior field of the simplest meaningful rotating body is described by the Kerr metric 
\cite{Kerr}. The Schwarzschild metric being the special case of the Kerr metric for spinless  bodies. 

On a previous work \cite{Steklain}, the Hill Problem is study in the framework of a pseudo-Newtonian potential that mimics some properties 
of the Schwarzschild metric, the Paczy\'nski-Wiita potential \cite{Paczynski}
\begin{equation}
\label{PW}
\Phi _{PW}= -\frac{GM}{r-r_g},
\end{equation}
 where $G$ is the gravitational constant, $M$ is the mass of the central body and $r_g =GM/c^2$ is the  Schwarzschild or  gravitational 
radius ($c$ is the speed of light). The aim of this work is to study the Hill problem taking into account the spin 
angular momentum of the central 
body, with a potential that mimics the dynamics of the Kerr metric. There are several potentials, 
originally used to study accretion disks around black holes, that fulfill this requirement. Some of this 
potentials are the Smerak-Karas potential \cite{sk} and the Mukhopadhyay potential \cite{Mukhopadhyay}. 
However, the potential proposed by Artemova, Bj\"ornsson and Novikov  (ABN potential) \cite{Artemova} was 
chosen due to (i) its agreement with the last stable and marginally bound orbits, obtained from the Kerr 
metric itself; (ii) its simple form, a natural extension of the Paczy\'nski-Wiita potential.

This work can be considered as a ``zeroth order'' approach to a general relativistic Hill problem with a spinning central body. Due to the 
complexity of the Einstein equations and its rigorous approximations, like the post-Newtonian expansions, it is worth to begin with this 
simplified approach to have an idea of the size of the quantities involved. These results can serve as a starting point for a more complete 
treatment of the problem. 

We shall compare the stability of orbits of the third body in the pseudo-Newtonian general relativistic simulation for different spins of
the central body with parameters that are typical for a system formed by a supercluster, a galaxy and a star. In this system the 
influence of the spinning on the dynamics can be easily seen.   

\section{The ABN Potential}

Artemova, Bj\"ornsson and Novikov \cite{Artemova} proposed two potentials to use in modeling accretion disks around a rotating black 
hole. They demanded that their potentials should have three properties: (i) the free-fall acceleration should have the form analogous to 
the Paczy\'nski-Wiita potential (equation (\ref{PW})); (ii) the free-fall acceleration must tend to infinity when tends to the event 
horizon of the black hole; and (iii) the position of the extremum of the boundary condition function must coincide with the position of the
last stable circular orbit in the exact relativistic problem. The boundary condition function reads
\begin{equation}
f(r)=1-\frac{l_{in}}{l(r)},
\end{equation}
 where $l(r)=r^2\Omega (r)$  is the Keplerian specific angular momentum at radius $r$ 
( $\Omega (r)$ is the  angular velocity being $\Omega (r)=(GM/r^2)^{1/2}$ for the 
Newtonian gravitational potential). The constant $l_{in}$ is given by $l_{in} = l(r_{in})$, where $r_{in}$ is 
the radius of the last stable orbit  \cite{Artemova}. The first of the two the free-fall accelerations obtained is of the 
form
\begin{equation}
\label{art}
F_{ABN} = -\frac{GM}{r^{2-\beta}(r-r_1)^\beta},
\end{equation}
or, in the potential form,
\begin{equation}
\label{potart}
\Phi _{ABN} = -\frac{GM}{(\beta -1)r_1} \left[ \left( \frac{r}{r-r_1} \right) ^{\beta -1} -1 \right], 
\end{equation}   
where $r_1$ is the position of the event horizon. This position is determined from the angular momentum $a$ by the exact expression from 
general relativity,  
\begin{equation}
r_1=[1+(1-a^2)^{1/2}]r_g.
\end{equation}
The value of $\beta$ is given from the following equation:
\begin{equation}
\beta=\frac{r_{in}}{r_1}-1
\end{equation}
Where $r_{in}$ is again the last stable orbit. The exact position of this orbit is obtained from the general relativity (see Novikov and Frolov 
\cite{Novikov}, equation 4.5.12),
\begin{eqnarray}
\label{rin}
r_{in}=[3+Z_2\mp [(3-Z_1)(3+Z_1+2Z_2)]^{1/2}]r_g,
\\
Z_1=1+(1-a^2)^{1/3}[(1+a)^{1/3}+(1-a)^{1/3}],
\\
Z_2=(3a^2+Z_1^2)^{1/2}
\end{eqnarray}
where the upper and the lower signs of the equation (\ref{rin}) are for co-rotation and counter-rotation respectively\footnote{In Ref.
\cite{Artemova}  equation (\ref{rin}) appears only with the minus sign, but the correct form is the above, as in Novikov 
and Frolov \cite{Novikov}.}. Note that the parameters $r_{in}$ and $\beta$ depend only on the angular momentum $a$. 

The radius of the last stable orbit, $r_s$ matches exactly with the last stable orbit in Kerr geometry, as demanded on the construction of 
this potential. In Table \ref{tabxb} the values of the marginally bound orbit, $r_{mb}$ are listed for the potential above and the Kerr 
geometry for various values of $a$. Note that they are in a good agreement with each other. Mukhopadhyay \cite{Mukhopadhyay} has pointed
that using the ABN potential for negative values the error in $r_b$ may be upto 500\%. However, if the
correct equation for counter-rotation is used (equation (\ref{rin}) with the lower signal), the error in $r_b$ is less than 2\%.

\begin{table}
\caption{\label{tabxb}Values of $r_{mb}/r_g$ for the ABN potential and Kerr geometry for various values of $a$}
\centering
\begin{tabular}{c c c c c c c}
\hline\hline
$a$ & 0 & 0.1 & 0.3 & 0.5 & 0.7 & 0.998 \\
\hline
$\Phi _{ABN}$ & 4 & 3.797 & 3.370 & 2.904 & 2.376 & 1.069 \\
\hline
Kerr & 4 & 3.797 & 3.373 & 2.914 & 2.395 & 1.091 \\
\hline\hline
$a$ & 0 & -0.1 & -0.3 & -0.5 & -0.7 & -0.998 \\
\hline
$\Phi _{ABN}$ & 4 & 4.198 & 4.578 & 4.942 & 5.288 & 5.746 \\
\hline
Kerr & 4 & 4.198 & 4.580 & 4.949 & 5.308 & 5.825 \\
\hline
\end{tabular}
\end{table}
      
For the second potential a different expression for the free-fall acceleration $F_{ABN}$ is obtained. We shall use (\ref{potart}) because
it is singular in the event horizon $r=r_1$, like the Paczy\'nski-Wiita potential. The second potential proposed by ABN does not have the 
above mentioned property.

\section{Modified Hill Problem}

\begin{figure}
\resizebox{\hsize}{!}{\includegraphics{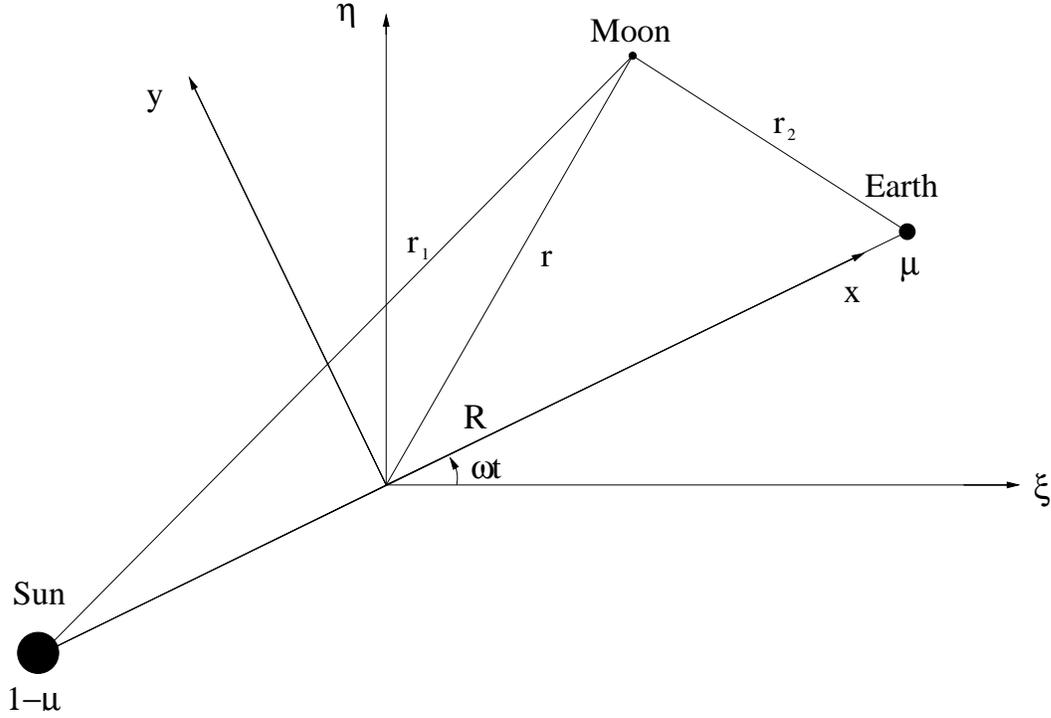}}
\caption{The planar, circular restricted three-body problem, on a inertial coordinate system $(\xi,\eta)$ and on a rotating coordinate
 system $(x,y)$. Note that on the rotating coordinate system the positions of the two massive bodies are fixed.}
\label{syst}
\end{figure}

The Hill problem is a special case of the circular, planar restricted three-body problem, as mentioned before. In this problem there is a 
system of two massive bodies with masses $m_1$ and $m_2$, in circular orbits around their center of mass and a third massless body moving
under influence of this system without perturbing it. In the Hill problem the body with mass $m_1$ is such that $m_1 \gg m_2$ and is far
away of the system, so it constitutes just a perturbation for the two-body system formed by the body with mass $m_2$ and the massless body.
We can choose units of mass such that $G(m_1+m_2)=1$. In this way we can take the units of mass of the two massive bodies respectively as 
$1-\mu$ and $\mu$, where $\mu=Gm_2$. We take units of distance such that the distance between the two massive bodies is $R=1$ and units of 
time such that the angular velocity of the rotating frame in which the two massive bodies are fixed is $\omega =1$ (see Fig. \ref{syst}).  

After placing the origin of the coordinate system in the body of mass $\mu$ and considering the motion only in a disk of radius 
$\mu^{1/3}$, we obtain the Newtonian Hill equations  \cite{Stuchi},
\begin{eqnarray}
\label{newtonhillx}
&&\ddot x = 2\dot y +3x - \frac{x}{r^3},
\\
\label{newtonhilly}
&&\ddot y = -2\dot x - \frac{y}{r^3},
\end{eqnarray}
where $r=\sqrt{x^2+y^2}$. Replacing the Newtonian potential by the ABN potential we obtain the modified Hill equations,
\begin{eqnarray}
\label{abnhill1}
\ddot{x}= 2\dot{y} + 3x -\frac{x}{r^{3-\beta}(r-r_1^*)^{\beta}},
\\
\label{abnhill2}
\ddot{y}=-2\dot{x}-\frac{y}{r^{3-\beta}(r-r_1^*)^{\beta}}.
\end{eqnarray}  
These equations can also be written as
\begin{eqnarray}
\ddot{x} - 2\dot{y} = -\frac{\partial U_{ABN}}{\partial x},
\\
\ddot{y} + 2\dot{x} = -\frac{\partial U_{ABN}}{\partial y},
\end{eqnarray}
where $U_{ABN}$ is the modified Hill potential 
\begin{equation}
\label{modifpot}
U_{ABN} = -\frac{3}{2}x^2  -\frac{\mathbf{1}}{(\beta -1)r_1^*} \left[ \left( \frac{r}{r-r_1^*} \right) ^{\beta -1} -1 \right] = -\frac{3}{2}x^2 + 
\Phi _{ABN} ,
\end{equation}
and $r_1^*=\mu^{-1/3}r_{1}$, in units such that the separation between the two massive bodies is  $R=1$, like in the Newtonian case. 

The modified Jacobi constant, in this case, is given by
\begin{equation}
C_{JABN} = \frac{1}{2}\left( \dot{x}^2+\dot{y}^2 \right) - \frac{3}{2}x^2 - 
\frac{\mathbf{1}}{(\beta -1)r_1^*} \left[ \left( \frac{r}{r-r_1^*} \right) ^{\beta -1} -1 \right].
\end{equation}

\section{Stability of orbits}

We shall use for the parameter $r_1^*$ the value $r_1^*= 5.10^{-6}$, that is a typical value for a system formed by a supercluster, a 
galaxy and a 
star. We compare the modified Newtonian systems with the angular momentum $a$ varying from $-0.5$ to $0.5$. 
The values of $\beta$ vary, respectively, from $3.05$ to $1.27$.     

\subsection{Fixed points analysis \label{subsec:fixedpoints}}

The Newtonian Hill problem has two well-known fixed points, the Lagrangian points $L_1$ and $L_2$. These 
points are of  saddle-center type, and are located at the positions $(\pm \sqrt[3]{1/3},0)$. For 
the ABN potential it can be guessed that new fixed points, in particular saddle points, arise due to the 
$r-r_1^*$ dependence of the denominator. However, if the parameter $r_1^*$ is small, the nature of the fixed points remains
 unchanged. The equation for the $x$ component of the fixed points (obtained by setting $\dot{x}$, 
$\dot{y}$, $\ddot{x}$, $\ddot{y}$ equal to zero in equations (\ref{abnhill1}) and (\ref{abnhill2}) and 
replacing $y=0$) is,
\begin{equation}
\label{eqfp}
|x|^{3-\beta}\left( |x|-r_1^*\right) ^{\beta}=\frac{1}{3}.
\end{equation}

If $r_1^*$ is small compared to $x$ (in this work $r_1^*$ is of order $10^{-6}$ while
$x$ scales as unity) the equation (\ref{eqfp}) reads, approximately,
\begin{equation}
|x|^3 \left( 1-\beta \frac{r_1^*}{|x|} \right) \approx \frac{1}{3}.
\end{equation}

For the values of $C_J$ so that the system is bounded, the value of the $x$ 
component of the fixed points is
also bounded. 
 There are one real and two complex conjugate solutions for this approximate equation. For
  higher values of 
$r_1^*$  the  exact equation  \ref{eqfp}) can have  more than one solution, they may  possibly  lead to interesting results, but these  systems may lack of physical significance.

The Jacobian of the system calculated at the fixed points reads
\begin{equation}
J_F(x_f,y_f)= \left(
\begin{array}{cccc}
0 & 0 & 1 & 0
\\
0 & 0 & 0 & 1
\\
9+3\beta \frac{r_1^*}{|x_f|-r_1^*} & 0 & 0 & 2
\\
0 & -3 & -2 & 0
\end{array}
\right).
\end{equation}
The associated characteristic polynomial  is
\begin{equation}
\left(\lambda ^2 +3\right) \left(\lambda ^2 - 9-3\beta \frac{r_1^*}{|x_f|-r_1^*}\right) =0.
\end{equation}
The eigenvalues are
\begin{eqnarray}
\lambda=\pm i\sqrt{3}
\\
\lambda=\pm \sqrt{9+3\beta\frac{r_1^*}{|x_f|-r_1^*}}.
\end{eqnarray}
For $r_1^*$ small enough [ $r_1^* < 3|x_f|/(3+\beta)$] there are two eigenvalues purely imaginary and two 
real, so the fixed points are of type \emph{saddle-center}, just as in the Newtonian Hill problem.

\subsection{Poincar{\'e} sections}

The orbits in the Newtonian as well in the modified Newtonian Hill problem are the solutions of a four-dimensional dynamical system with 
variables $(x,y,\dot x,\dot y )$. Since we have an integral of motion, $C_{JABN}$, the motion is reduced to a  three-dimensional system, 
we can take $(x, \dot x , y)$ as independent variables.We shall study surface of section (Poincar\'e sections) evaluating the orbits for 
different values of the Jacobi constant and registering the crossings of the hypersurface $y=0$ with $\dot y > 0$. 

\begin{figure}
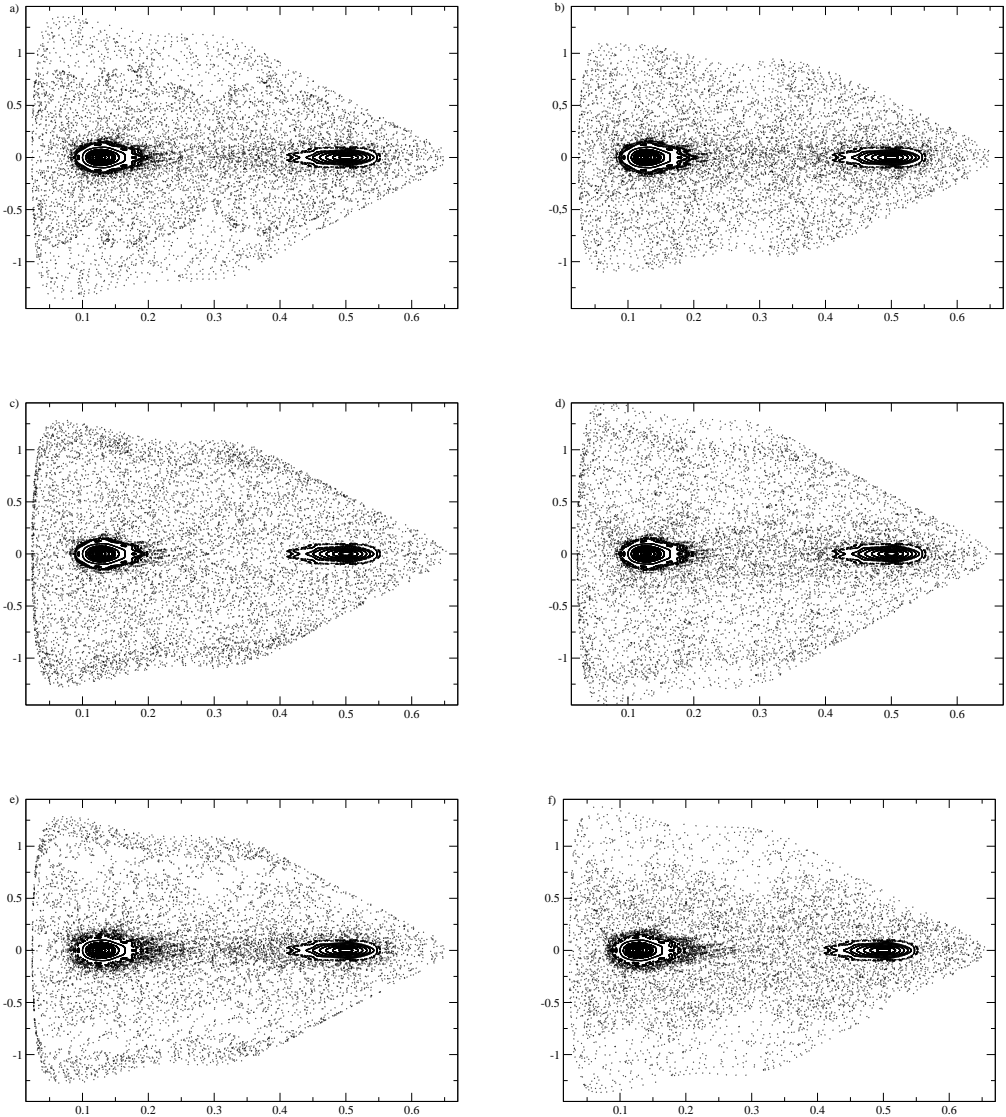
 
\includegraphics[totalheight=0.23\textheight,viewport=-20 0 690 558,clip]{secao1a}
\hfill
\includegraphics[totalheight=0.23\textheight,viewport=-20 0 690 558,clip]{secao3a}
\hfill
\includegraphics[totalheight=0.23\textheight,viewport=-20 0 690 558,clip]{secao5a}
\hfill
\includegraphics[totalheight=0.23\textheight,viewport=-20 0 690 558,clip]{secao-1a}
\hfill
\includegraphics[totalheight=0.23\textheight,viewport=-20 0 690 558,clip]{secao-3a}
\hfill
\includegraphics[totalheight=0.23\textheight,viewport=-20 0 690 558,clip]{secao-5a}
\hfill
\caption{Poincar\'e sections for different values of the angular momentum $a$. Positive values a) $a=0.1$ , b) $a=0.3$ and c) $a=0.5$  and 
negative values d) $a=-0.1$, e) $a=-0.3$ and f) $a=-0.5$. The rate of destruction of KAM tori is greater for the negative values. Note that, for 
larger values of $a$, new islands appear in the regular region, indicating transition to chaos.}
\label{poincarec}
\end{figure}

\begin{figure}
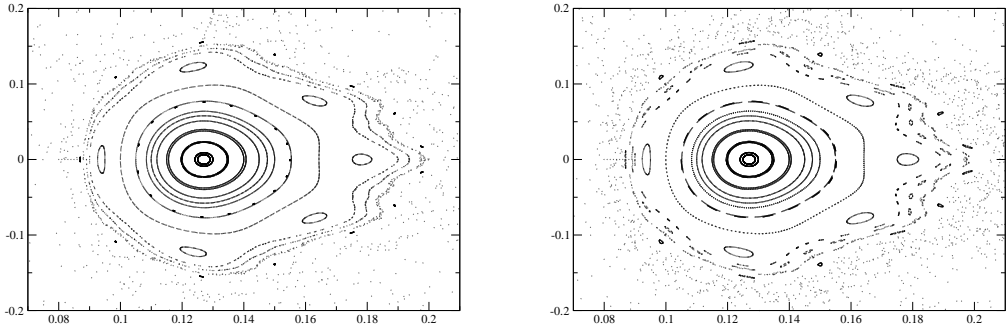

\includegraphics[totalheight=0.23\textheight,viewport=-20 0 690 558,clip]{zoom5b}
\hfill
\includegraphics[totalheight=0.23\textheight,viewport=-20 0 690 558,clip]{zoom-5b}
\caption{Magnifications of areas 
in the Poincar\'e sections, at left for angular momentum $a=0.5$ and at right for angular momentum $a=-0.5$. Note that for
$a=-0.5$ the tori are more affected by the perturbation.}
\label{zoompoinc}
\end{figure}

 The results for 
$C_J= -2.17$ (typical value for a bounded system) are shown on the Fig. \ref{poincarec}. In Fig. \ref{zoompoinc} there is a magnification of the right portion for the values $a=0.5$ and $a=-0.5$. We see that, as the angular momentum $a$ increases in modulus (from (a) to (c) and from
(d) to (f) in Fig. \ref{poincarec}) some  Kolmogorov-Arnold-Moser (KAM)  tori are deformed and others destroyed, indicating the transition of the system from regular to chaotic 
behaviour. This transition is more evident for the sequence of negative values of $a$, i.e., the
sequence from (d) to (f), so the destruction of KAM tori is faster for counter-rotation. It means that, for the counter-rotation case, a larger region of the phase space is chaotic when compared with 
the associated 
co-rotation system (same angular velocity $a$ in modulus). It does not means necessarily that the counter-rotating systems are more 
unstable
than the co-rotating ones, since the orbits are still bounded by KAM tori that are not destructed by the perturbation. Besides, the 
dependence on initial conditions, the main characteristic of chaotic systems, still must be analyzed by appropriate tools, like Lyapunov 
exponents. This analysis can decide if a system is more chaotic than other \cite{Ruelle}. 

\subsection{Lyapunov exponents}

To analyze quantitatively the orbits stability we study the Lyapunov exponents for the systems above described. The Lyapunov
characteristic number ($\lambda$) is defined as the double limit, 
\begin{equation}\label{lcn}
\lambda=\lim_{\scriptsize \begin{array}{l} \delta_0\rightarrow 0 
\\
t\rightarrow \infty 
\end{array}
}
 \left[\log(\delta /\delta _{0}) \over t\right] , 
\end{equation}
where $\delta _{0}$ and  $\delta $\ are the deviation of two nearby orbits at times $0$ \ and $t$\, respectively (see Alligood et al. 
\cite{Yorke}). We get the largest $\lambda$ using the technique suggested by Benettin et al. \cite{Galgani}  and  the algorithm of Wolf et al. \cite{Wolf}.

\begin{figure}
\resizebox{\hsize}{10cm}{\includegraphics[totalheight=0.2\textheight,viewport=-20 0 750 558,clip]{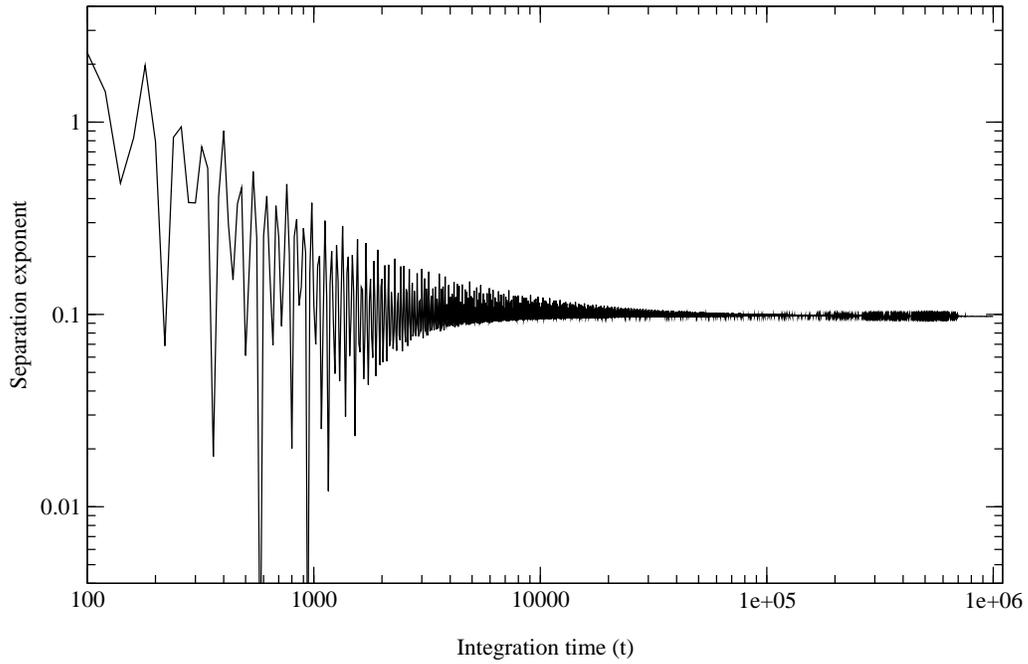}}
\caption{Time evolution of the separation exponent. It converges to the corresponding Lyapunov characteristic number ($\lambda$) 
for large $t$.} 
\label{evolucao}
\end{figure}

The Lyapunov exponents are not absolute, but dependent on the choice of the time scale. We recall that  we have fixed the time scale by 
the requirement that $\omega=1$.  This  defines a time unit that is natural to each particular system at it is  given in terms of the characteristic period. In this work the analysis is made by  varying only the angular momentum $a$, so the direct comparison between the different Lyapunov exponents is valid. Each coefficient was computed until  convergence is reached. To achieve this precision the system of equations  was integrated for at least one hundred
thousand periods, as shown in Fig. \ref{evolucao}.  The initial conditions used to perform the integration must be chosen in the bounded region of
the system. To estimate the limit of this region we use the Lagrangian points of the Newtonian Hill system, given numerically by $(\pm 0.69,0)$. For
safety we chose the initial position $(0.3,0)$. The initial velocity is obtained from the Jacobi constant $C_J=-2.17$, choosing $\dot{x}=0$.  As can be seen in the subsection \ref{subsec:fixedpoints} the eigenvalues obtained from the Jacobian of the system remain small, so the system is non-stiff. The calculations are performed with the aid of the Burlisch-Stoer method with step control, that works well for non-stiff systems, and the error due to the integration is proportional to the tolerance imposed   ($10^{-10}$). This relative error is kept small enough, so the errors of the coefficients are basically estimate from the fluctuations  at the end of  each evolution  (Fig. \ref{evolucao}). The absolute error, according to these fluctuations, is on order of $10^{-8}$. 

\begin{figure}
\resizebox{\hsize}{10cm}{\includegraphics[totalheight=0.2\textheight,viewport=-20 0 750 558,clip]{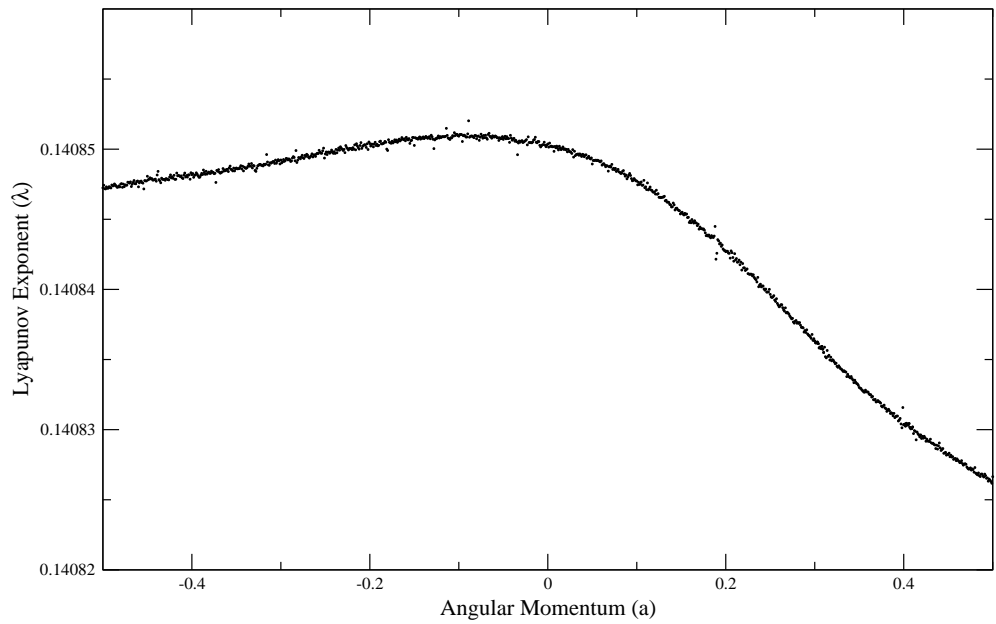}}
\caption{Lyapunov exponents for different values of $a$.} 
\label{lyapunov}
\end{figure}

The Lyapunov exponents obtained for this system are shown in Fig. \ref{lyapunov}. As can be seen from this figure, negatives value of $a$ are associated to a larger Lyapunov exponent 
when compared with their correspondent positive values. It means that two neighbour orbits separate from each other faster in the counter-rotating system than in the co-rotating system, meaning that the counter-rotating system is more unstable. Nevertheless, this dependence on parameter $a$ is very small for real systems. The reason is the fact that the apparent event
horizon is very small when compared with typical distances in the system, and the angular momentum $a$ has only little influence on this horizon as already pointed by Bardeen, Press 
and Teukolsky \cite{Bardeen}. 
%This small dependence of the Lyapunov exponents on the parameter $a$, in contrast with the strong dependence on position, 
%difficults a complete analogy with the Poincar\'e sections by comparing different initial
%conditions. Besides, in a small region one can show that indeed the counter-rotating system presents stonger dependence on initial 
%conditions, as shown in Fig. \ref{varr}. 
We shall study this dependence in a future work taking another model for the Hill problem closer 
to the exact general relativistic system.


%\begin{figure}
%\resizebox{\hsize}{10cm}{\includegraphics[totalheight=0.2\textheight,viewport=-20 0 750 558,clip]{varr}}
%\caption{Lyapunov exponents for different initial conditions, for $a=0.5$ and $a=-0.5$.} 
%\label{varr}
%\end{figure}

\subsection*{d) Fractal escape and fractal dimension}

The Poincar\'e sections   and Lyapunov exponents  were obtained for values of the Jacobi constant such that the systems are  bounded. For values larger than 
$C_{Jbounded}= -2.16$ the systems  are unbounded and the third body can escape by two different routes \cite{Steklain}. 
For open systems that have more than one route to  escape we can apply the fractal escape technique used by  Moura and Letelier 
\cite{Moura} in the study of the classical H\'enon-Heiles problem.  In this method the basins of the escape routes are obtained for a set
of initial conditions.  For chaotic systems, we  have the existence of fractal basin boundaries (FBB)  indicating a great instability of 
the orbits. In our case we chose a subset of the accessible  phase space at a fixed Jacobi constant, defined by a segment $|x|\leq a$, 
$y=b$ and $0 \leq \theta \leq   2\pi$, where $a$ and $b$ are constants to be chosen appropriately and $\theta$ is the angle that defines 
the direction of the  velocity with respect the $x$-axis. Then the trajectories are  integrated numerically, we have  three different 
cases: (i) the body escapes to $x \rightarrow +\infty$, (ii) it  escapes to $x \rightarrow -\infty$, and (iii) the particle does  not scape
during the integration time. We take  the  integration time long enough to  assure that our results be consistent.

To show the difference between  systems with different values of the parameter $a$, we calculate the dimension of the basin boundaries 
obtained for the different  systems. The dimension used is the  box-counting dimension that can be easily obtained, see for instance,  
Ott \cite{Ott} and  Grebogi et al. \cite{Grebogi}.
If we displace a determined point of a basin to another on a distance $\epsilon$, the probability that this new initial condition does not 
belong to the same basin of the old one is, for small $\epsilon$, $P(\epsilon)\propto \epsilon^{D-d}$, where $D$ is the dimension of the set
($2$ in our case) and $d$ is the box counting dimension, also called exterior or fractal dimension when not integer. In order to calculate 
this fractal dimension, for several values of $\epsilon$, we displace the $x$ coordinate of all the points from one of the basins 
($x \rightarrow +\infty$), and count the number of points that does not  belong to the same basin. Then we compute the   the fraction of
 numbers that does not  belong to the same basin, $P(\epsilon)$. We plot $\ln P(\epsilon)$  in function of $\ln\epsilon$.  The inclination 
of the straight line gives us $D-d$. 

\begin{figure}
\resizebox{\hsize}{10cm}{\includegraphics[totalheight=0.2\textheight,viewport=-20 0 690 558,clip]{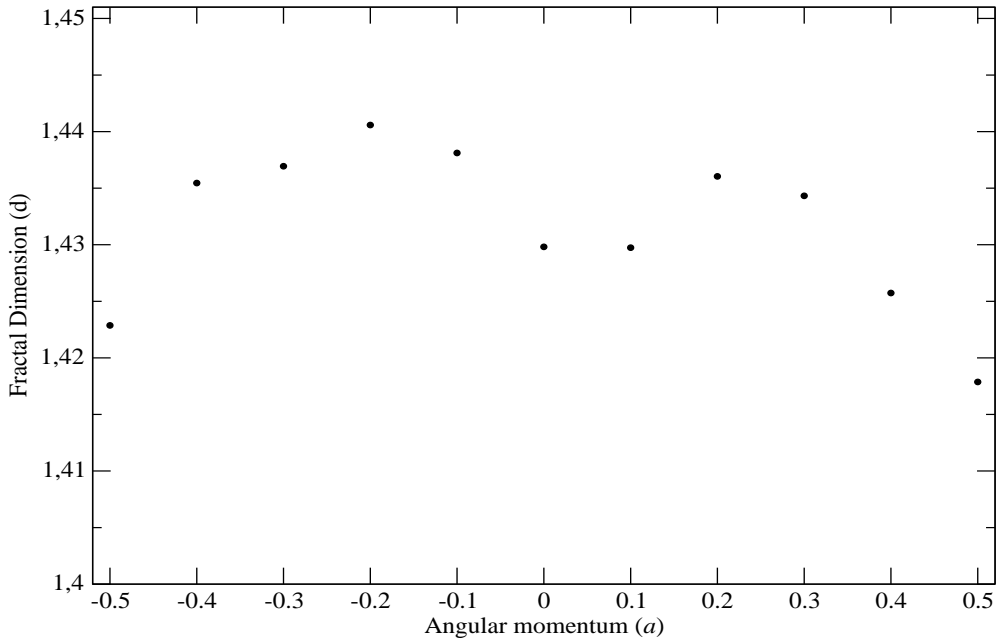}}
\caption{Fractal dimensions for different values of $a$.} 
\label{dimfrac}
\end{figure}

 The values for the fractal dimension obtained are shown in the Fig. \ref{dimfrac}, with error of order $3.10^{-3}$. This uncertainty is
mainly due to the error in the computation of the line slope. It can be seen that the fractal dimension for
negative values of $a$ are larger than the correspondent positive values. The fractal dimension for unbounded systems have a close
relationship with the chaoticity of the systems \cite{Steklain},\cite{Moura},  larger fractal dimensions are related to systems 
that are more unstable. In this sense we can conclude that counter-rotating systems are more unstable than their correspondent co-rotating 
cases. 
% This is a indirect result, only valid for unbounded systems.
The validity of this result is limited, as it is applicable only to unbounded systems and the error in calculations are of the order of
the variation of data. For bounded systems the analysis of the Lyapunov exponents 
in the previous section confirms the validity of this result. The values of $d$ are computed only for a few values of $a$. Is possible that this
small density of points can hide a more complex behaviour for this system. Unfortunately to achieve the same density of points used for the
Lyapunov exponents is still prohibitive due to the time taken to perform the calculations. We shall improve this method in our future works.

\section{Conclusions}

In this work we show that we can simulate the dragging of inertial frames by a pseudo-Newtonian potential. We also show, by Poincar\'e 
maps,  Lyapunov exponents and fractal escape techniques, that there is a dependence of the stability of the orbits on the spin angular
momentum of the 
central body. The bounded and unbounded systems where the movement of particle around the central body is opposite to its spin (counter-rotating) are more unstable than systems where the two rotations are in the same direction (co-rotating). This preliminary result is in accord with previous studies of the stability of orbits of particles moving around 
spinning centers of attraction \cite{leteliervieira},\cite{leteliergueron}.  This effect is small when compared with the influence of the position of the event horizon \cite{Steklain}. Otherwise, it can have a larger influence on the Lyapunov exponents.  In a future work this
feature will be studied in detail to compare the chaoticity of co-rotating and counter-rotating orbits in a full relativistic system, as
geodesics in a Kerr black hole with halos \cite{leteliervieira} or multipolar deformations \cite{leteliergueron}.

\section{Acknowledgements}
We want to thank CNPq and FAPESP for partial financial support.

\end{document}